\documentclass[english,pre,showpacs,floatfix,preprint]
{revtex4}
\usepackage[T1]{fontenc}
\usepackage[latin1]{inputenc}
\usepackage{babel}
\usepackage{epsfig}
\usepackage{amsmath}
\usepackage{amssymb}
\usepackage{subfigure}
\usepackage{graphicx}
\usepackage{psfrag}

\newcommand{\p}[2]{\frac{\partial\, #1}{\partial\, #2}\,}

\newcommand{\Z}{\mathbb{Z}}

\newcommand{\V}[1]{\mathbf{#1}}

\newcommand{\eq}[1]{$\mathrm{Eq.}$~\eqref{#1}}
\newcommand{\figref}[1]{$\mathrm{Fig.}$~\ref{#1}}
\newcommand{\secref}[1]{$\mathrm{Sec.}$~\ref{#1}}

\newcommand{\av}[1]{\langle\, #1 \,\rangle}

\begin{document}

\title
 {\bf Resonance-like phenomena of the mobility of a chain of nonlinear coupled oscillators in a two-dimensional periodic potential}
\author{S. Martens, D. Hennig, S. Fugmann and L. Schimansky-Geier}
\medskip
\medskip
\medskip
\affiliation{
Institut f\"{u}r Physik, Humboldt-Universit\"{a}t Berlin,\\Newtonstr. 15, 12489 Berlin, Germany}

\begin{abstract}

\noindent We study the Langevin dynamics of a two-dimensional discrete oscillator chain absorbed on a periodic substrate and subjected to an external localized point force. Going beyond the commonly used harmonic bead-spring model, we consider a nonlinear Morse interaction between the next-nearest-neighbors. We focus interest on the activation of directed motion instigated by thermal fluctuations and the localized point force. In this context the local transition states are identified and the corresponding activation energies are calculated. As a novel feature it is found that the transport of the chain in point force direction is determined by stepwise escapes of a single unit or segments of the chain due to the existence of multiple locally stable attractors. The non-vanishing net current of the chain is quantitatively assessed by the value of the mobility of the center of mass. It turns out that the latter as a function of the ratio of the competing length scales of the system, that is the period of the substrate potential and the equilibrium distance between two chain units, shows a resonance behavior. More precisely there exist a set of optimal parameter values maximizing the mobility. Interestingly, the phenomenon of negative resistance is found, i.e. the mobility possesses a minimum at a finite value of the strength of the thermal fluctuations for a given overcritical external driving force.
\end{abstract}

\maketitle

\section{Introduction}
\noindent Transport phenomena play a fundamental role in many physical systems.  For systems that evolve in an external potential which possesses metastable states the thermally activated escape over potential barriers, as the precondition for transport, is the mostly studied situation. The related escape problem, often referred to as the Kramers problem \cite{Kramers}, has been reviewed e.g. in \cite{50yKramers}. Due to its ubiquity and simplicity a spatially periodic potential is employed in a number of applications including Josephson tunneling junctions \cite{machura07,machura07_2,Nagel08}, phase-locked loops \cite{Haken}, rotation of dipoles, charge-density wave \cite{Gruner}, dislocation \cite{mudimer3},
diffusion of atoms and molecules on crystal surfaces \cite{CPD1}, biophysical processes such as neural activity and intracellular transport \cite{Schimansky1,Zaks03,Risken}. Exact expressions for the characteristic quantities of motion of one single Brownian particle like the net-current and the diffusion coefficient are given in \cite{Schimansky1}.

 In the last decades, the interest in the theory of Brownian motion of interacting particles \cite{Vollmer,Schneider,Braun90,Difftilted} has grown in several fields of science. In this context the study of the diffusion process \cite{Pijper,Lu} and the mobility of strongly interacting atoms subjected to a periodic potential and driven by an external force is a first step towards the understanding of solid friction at the atomic level \cite{Brushan,Braun}. Recently some studies have considered the transport of dimers in a one-dimensional ($1$D) washboard potential under the impact of spatially \textit{uniform} dc and ac forces \cite{mudimer3, Braun03,Fusco03,Goncalves05,Heinsula,Hennig08} which are applied to all particles. A complicated non-monotonous behavior of the mobility depending both on the external driving and on the ratio between the competing length scales of the system is found. The latter plays the role of an internal degree of freedom which generates a ratchet effect \cite{Haeussler97,Juelicher97,Wang04,Wang05,Menche06,Wang07,vGehlen08}.

 Particularly in biophysical contexts, the extension to coupled multi-dimensional systems, e.g. the transport of long and flexible polymers across membranes \cite{Sebastian00_61,Sebastian00_62,Debnath07} or DNA electrophoresis \cite{Park,Huopaniemi}, has recently attracted considerable interest. Motivated by experimental advances in the manipulation and visualization of single polymers using optical
\cite{Ashkin} and magnetic \cite{Strick} tweezers or scanning force microscopy \cite{Rabe} the external driving acting on the system can also be modeled by an external point force which is applied at one single unit \cite{Hennig04,Kraikivski,Santo02}.

In this paper we consider the noise assisted transport of a two-dimensional ($2$D) discrete oscillator chain confined onto a periodic substrate and subjected to an external \textit{localized} point force. Our theoretical study is related to single-molecule experiments using scanning force microscopes. In the commonly used bead-spring model the next-nearest neighbors are coupled harmonically. This assumption is valid only for small elongation from the equilibrium distance. In order to take the nonlinear character of the coupling into account, we introduce an interaction potential of Morse-type which allows bond rupture \cite{Puthur02,Dudko06}. Due to the imposed next-nearest neighbors coupling and the fact that the Morse interaction potential is rotationally symmetric all configurations with the same distribution of distances between two coupled units possess the same amount of energy independent of the distribution of the angles between two coupled constituents of the chain. Therefore the bending rigidity $\kappa$ and the persistence length $L_p=\kappa/T$, respectively, are equal to zero. Since the chain is stretched during the motion as a result of applied external forces and the substrate potential force we can neglect excluded volume interactions.

 In our present study we focus interest on the impact of the ratio between the competing length scales of the system and the external driving on the directed motion of the chain. The latter is instigated by thermal fluctuations and quantitatively assessed by the mobility. The main question is how both the spatially localized driving and the nonlinear interaction potential between the many degrees of freedom influence the mobility of the chain? More precisely, under which conditions differs the mobility of the considered system strongly from the one for $1$D dimer \cite{Heinsula,mudimer3} and does there exit sets of parameters for which the latter coincide?

This paper is organized as follows: In \secref{sec:model} we
introduce the model. The transition state configurations are identified in
\secref{sec:TST}. Further, in \secref{sec:eact} we derive the
scaling behavior of the activation energy in the limit of weak coupling. The parameter values influence on the mobility is studied in \secref{sec:mob}. We conclude with a summary and discussion of our results.

\section{The model} \label{sec:model}
\noindent We study a two-dimensional nonlinear coupled oscillator chain consisting of $N$ particles of equal mass $m$ evolving in a $2$D substrate under the influence of an external dc point force with magnitude $F$. The point force is applied at one single constituent at site $n_0$. The coordinates of the $n$-th unit,$n=1,\ldots,N$, in the x-y-plane are given by $\vec{q}_n(t)=(x_n(t) , y_n(t))^T$.

\begin{figure}
\includegraphics[width=0.5\textwidth]{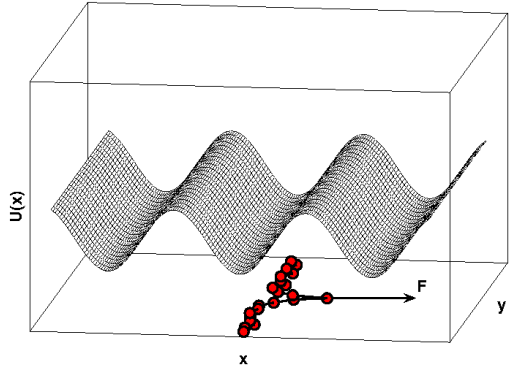}
\caption{(Color online) Schematic view on the potential landscape with a segment of the chain inside. The parameter values are $A=1$ and $L=2\pi$. \label{fig:on-site}}
\end{figure}

The interaction of the particles with the substrate is modeled by the on-site potential
\begin{align}
 U(x_n)=&\frac{A}{2}\,\left(1-\cos\left(\frac{2\pi}{L}x_n\right)\right)\,, \label{eq:onsite}
\end{align}
with periodicity $L$, i.e. $U(x+L)=U(x)$, and potential height A. Note that the on-site potential is translational invariant in the y-direction. The dc point force acting on the unit at site $n_0$ is introduced by the additional potential term
\begin{equation}
-F\,x_n\,\delta_{n,n_0}\,.
\end{equation}
A segment of the two-dimensional periodic on-site potential $U(x_n)$ and the position of the chain is depicted in \figref{fig:on-site}.

Each particle is connected to its two next-nearest-neighbors by nonlinear springs described by the Morse potential \cite{Morse}
\begin{align}
    W(r_{n+1,n})=D\,\left[1-\exp\left(-\alpha\,\left(r_{n+1,n}-l_0\right)\right)\right]^2\,. \label{eq:morse}
\end{align}
The Euclidean distance between two units at site $n$ and $n+1$ is identified with $r_{n+1,n}=\sqrt{(x_{n+1}-x_{n})^2+(y_{n+1}-y_{n})^2}$ and the parameter $l_0$ denotes their equilibrium distance. The dissociation energy of a bond and the inverse range of the potential are determined by $D$ and $\alpha$, respectively.

\begin{figure}
\psfrag{l}{\footnotesize $\mathbf{l_0}$}
\includegraphics[width=0.5\textwidth]{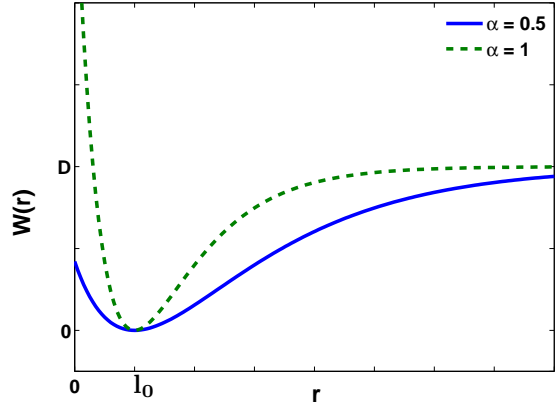}
\caption{(Color online) Morse potential $W(r)$ for various inverse interaction range $\alpha$. The parameter values are $D=10$ and $l_0=1$. \label{fig:morse}}
\end{figure}

In the overdamped limit the inertia is omitted and the dynamics at finite temperature $T$ is described by the Langevin equation (LE)
\begin{small}
\begin{align}
 \gamma\,\dot{\vec{q}}_n=\,-\frac{1}{m}\nabla_{\vec{q}_n}\lbrace& W(r_{n,n-1})+W(r_{n+1,n})+U(x_n)-F\,x_n\,\delta_{n,n_0}\rbrace+\vec{\xi}_n(t)\,. \label{eq:dglun}
\end{align}
\end{small}
Here $\gamma$ is the viscous friction coefficient per unit mass.

For convenient rescaling we introduce suitable space $q_{\mathrm{sc}}=L/2\pi$, energy $E_\mathrm{sc}=A$, and time units $t_{\mathrm{sc}}=m\gamma\,q_{\mathrm{sc}}^2/E_\mathrm{sc}$, respectively, and define the dimensionless quantities
\begin{align}
    \vec{\overline{q}}=&\,\frac{\vec{q}}{q_{\mathrm{sc}}}\,, & \overline{t}=&\,\frac{t}{t_{\mathrm{sc}}}\,, &  \overline{T}=&\,\frac{k_B\,T}{E_\mathrm{sc}}\,, & \vec{\overline{\xi}}(\overline{t})=&\,\sqrt{t_{\mathrm{sc}}}\,\vec{\xi}(t)\,, \notag \\
    \overline{F}=&\,\frac{q_\mathrm{sc}}{E_\mathrm{sc}}\,F\,, &
    \overline{D}=&\,\frac{D}{E_\mathrm{sc}}\,, & \overline{\alpha}=&\alpha\,q_{\mathrm{sc}}\,, & \overline{l}=&\frac{l_0}{q_{\mathrm{sc}}}\,.
\end{align}
Note that the new quantity $\overline{l}$ determines the ratio between the competing length scales of the system $l_0/L$. Below, we refer to $\overline{l}$ as the bond length and we omit the overbar in our notation.  Finally the dimensionless equations of motion read as
\begin{subequations}
  \begin{align}
  \dot{x}_n=\,&-\p{W(r_{n,n-1})}{r_{n,n-1}}\p{r_{n,n-1}}{x_n}-\p{W(r_{n+1,n})}{r_{n+1,n}}\p{r_{n+1,n}}{x_n}- \notag \\
  &-\frac{1}{2}\,\sin\left(x_n\right)+F\,\delta_{n,n_0}+\xi_n^x(t)\,, \label{eq:dglx} \\
  \dot{y}_n=\,&-\p{W(r_{n,n-1})}{r_{n,n-1}}\p{r_{n,n-1}}{y_n}-\p{W(r_{n+1,n})}{r_{n+1,n}}\p{r_{n+1,n}}{y_n}+ \notag \\ &\,+\xi_n^y(t)\,.
  \label{eq:dgly}
  \end{align} \label{eq:dgl}
\end{subequations}

The stochastic force $\vec{\xi}_n(t)$ represents random fluctuations stemming from the influence of the environment. The latter is Gaussian white noise with zero mean, $\av{\vec{\xi}_n(t)}=0$, and autocorrelation function $\av{\xi_i^{x,y}(t)\,\xi_j^{x,y}(s)}=2\,T\,\delta_{i,j}\delta_{x,y}\delta(t-s)$.

Throughout this work we impose open boundary condition (OBC), i.e. $x_1-x_0=x_{N+1}-x_N=0$ and $y_1-y_0=y_{N+1}-y_N=0$, and we use an odd number of units $N$. For the sake of symmetry we fix the pulled particle as the one situated at the center of the chain, i.e. $n_0=(N+1)/2$. Without loss of generality we set $\gamma,A$ and $m$ equal to $1$ and $L=2\pi$ for the periodicity. Consequently the scaling parameters equal $q_{\mathrm{sc}}=E_\mathrm{sc}=t_{\mathrm{sc}}=1$.

We remark that the metastable ($k$ even) and unstable ($k$ odd) states of the on-site potential, $\partial U(x_n)/\partial x_n=0$, are located at
\begin{align}
 x_U^k=\,(-1)^k\arcsin\left(2F\right)+k\,\pi\,, \label{eq:stateU}
\end{align}
as a result of the external point force. From this follows that in the limit of vanishing coupling the pulled oscillator cannot be trapped by the on-site potential given in \eq{eq:onsite} under any circumstances for a tilt larger than the critical value $F_{\mathrm{cr}}=1/2$. In order to characterize the relative strength of the interaction potential versus the on-site potential, we introduce the coupling strength $K$
\begin{align}
 K=\,\frac{W''(r)|_{r=l}}{U''(x_n)|_{x_n=0}}=\,4D\,\alpha^2\,. \label{eq:Kdef}
\end{align}
Note that $K$ depends quadratically on the inverse interaction range $\alpha$ and linearly on the dissociation energy $D$.

\section{Transition state}\label{sec:TST}
\noindent In the following we focus our interest on the escape
dynamics of the $2$D coupled nonlinear oscillator chain from a
metastable state over an energy barrier of the corresponding energy
hypersurface. This progress requires the activation energy $E_\mathrm{act}$ which coincides with the height of the energy barrier. We identify the transition state and study the
dependence of the associated activation energy on the system
parameters.

According to the classical transition state theory
\cite{Laidler,50yKramers}, transition states are specials
points in the $2N$-dimensional phase space. More precisely, a
transition state $\{\V{q}^\dagger\}$ is a hyperbolic fixed point obtained from the stationary system of
the deterministic dynamical system given by
\eq{eq:dgl} for
$\{\dot{x}_n\}=\{\dot{y}_n\}=0$. The corresponding Jacobian
possesses at least one real positive eigenvalues which corresponds to the movement along the reaction coordinate. All other eigenvalues are negative. We apply a multi-dimensional root finding algorithm using
the Newton-Raphson method to calculate the transition state from the corresponding stationary system.

As a result of the considered overdamped dynamics of the system and the reflection of the periodicity of the on-site potential $U(x_n)$ by the $2N$-dimensional phase space instead of one global basin of attraction, referred to as the running state, multiple locally stable attractors exist  \cite{50yKramers,Risken}. Therefore the phase space possesses many different transition states which can be either linked with one required unique activation energy or different ones. Since the escape rate
over an energy barrier $E_\mathrm{act}$ is assumed to be described by the Van't
Hoff-Arrhenius law \cite{50yKramers,Langer}, $r_\mathrm{esc} \propto \exp(-E_\mathrm{act}/T)$, we restrict our analytical consideration of possible escape processes to scenarios with low amount of activation energy.

The energy of one configuration is determined by the energy
functional $V(\{\V{q}\})$
\begin{align}
V(\{\V{q}\})=\,\sum\limits_{i=1}^N
U(x_i)+\sum\limits_{i=1}^{N-1}W(r_{i,i+1})-F\,x_{n_0}\,.
\label{eq:ensurf}
\end{align}
Further, the activation energy $E_{\mathrm{act}}$ is defined as the
difference between the energy of the transition state configuration
$\{\V{q}^\dagger\}$ and the energy of the chain at the initial
metastable state $\{\V{q}^\mathrm{min}\}$
\begin{align}
    E_{\mathrm{act}}=\,V(\{\V{q}^\dagger\})-V\left(\{\V{q}^\mathrm{min}\}\right)\,.
    \label{eq:eact}
\end{align}

\begin{figure}
\subfigure[Transition state profiles for weak coupling \label{fig:profa}]{ \includegraphics[width=0.5\textwidth]{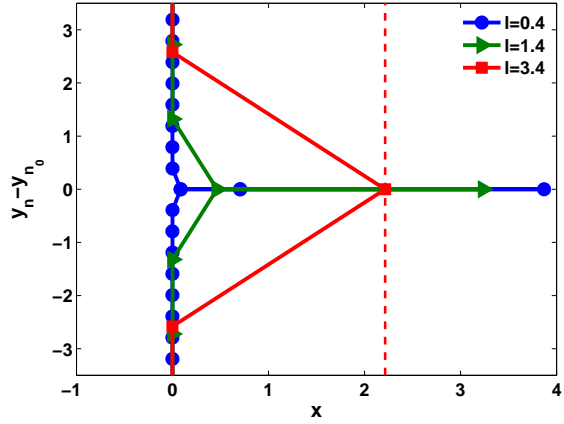}}
\subfigure[Transition state profiles for strong coupling \label{fig:profb}]{
\includegraphics[width=0.5\textwidth]{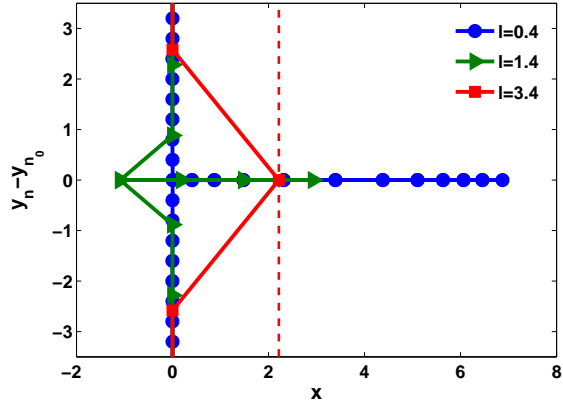}}
\caption{(Color online) Profile of the transition state belonging to the $1$st escape process for different values of bond length $l$ and of inverse range parameters $\alpha$. Only a
segment of the chain is shown. The dashed line represents the
position of the energy barrier $x_U^1$ of the pulled particle. The remaining parameter
values are $N=99,n_0=50,D=10,F/F_\mathrm{cr}=0.8,\alpha=0.1$ (a), and $\alpha=0.4$ (b). The values of the coupling strength are $K=0.4$ (a) and $K=6.4$ (b).}  \label{fig:prof}
\end{figure}

Below, we discuss the numerically calculated results for the transition state configuration (TSC) of the initial escape scenario in which only the pulled particle escapes from its initial minima to the next well of the on-site potential in positive x-direction. A sketch of this first
escape scenario is presented in \figref{fig:escsk} and is labeled by
$(1)$.  The numerically calculated TSC are depicted in \figref{fig:prof} for various values
of the bond length $l$ and for two different coupling strengths $K$.

At first glance, one can see that for $l \geq x_U^1$, see \eq{eq:stateU}, the
escape process is governed by an individual escape of the pulled
unit independent of the coupling strength. At the TSC the $n_0-$th
oscillator is always placed at the top of the energy barrier located
at $x_U^1$ while all other units rest at
the minimum of the on-site potential $U(x_n)$ without changes of
the respective bond length $l$.

For $l<x_U^1$ and all values of the coupling strength $K$, the pulled particle is placed beyond the energy
barrier of the on-site potential and its neighbors are elongated in
x-direction. According to the deterministic part of the equations of motion \eq{eq:dgly}, the
condition $\{\dot{y}_n\}=0$ is fulfilled either if the distance
between two neighboring particles equals the bond length $l$ or the
displacement $y_{n+1}-y_{n}$ is equal to zero. Since the adopted distance between two
neighboring units at the TSC differs from the bond length due to the
interaction with the substrate $U'(x_n)$, the obtained
structures resemble the shape of very thin needles, i.e.
$y_{n_0\pm i}-y_{n_0}=0$. The observed reflection symmetry in y-direction,
$|y_{n_0+i}-y_{n_0}|=|y_{n_0-i}-y_{n_0}|$, results from the choice of
$n_0$ at the central site of the chain. Note that as a consequence
of the imposed open boundary conditions all units of the chain are
elongated from their starting equilibrium positions at the transition
state configuration.

Comparing the obtained TSC for weak coupling in \figref{fig:profa}
and the ones for stronger coupling in \figref{fig:profb}, one
recognizes that with increasing coupling the number of elongated
units involved in the transition state grows. In the case of very
weak coupling, the units of the chain tend to diffuse via individual
steps. Despite that there result large elongations, $r_{n,n+1}\gg l$, during the
process the units remain bound to each other. With
further increasing value of coupling $K$  the oscillators move like
a rigid unit with $r_{n,n+1}\simeq\,l$ and thus display organized
collective behavior reflected in synchronized escape.

Such a dependence of the diffusion type on the coupling strength is already experimentally known \cite{CPD1,CPD2} and theoretical investigated \cite{Pijper,Lu,Romero04} for dimers diffusing on a surface.

Nevertheless, due to the considered overdamped dynamics of the chain multiple local domains of attraction exist for $F/F_\mathrm{cr}<1$. Hence, the transport of the chain in point force direction is determined by \textit{stepwise escapes} of single units or segments of the chain. These stepwise crossings are connected with configurational changes of the chain.

\section{Truncated trimer model and scaling behavior of the activation energy}\label{sec:eact}
\noindent In the following, we study separately the scaling behavior of the activation energy for different escape steps in order to assess the time scales of the latter. The characteristic time one particle needs to escape from $x_0 \to x_0+2\pi$ ($x_0$ is one arbitrary reference point) is determined by the first moment of the first escape time distribution $T_\mathrm{esc} =\av{t(x_0 \to x_0+2\pi)} \propto \exp (E_\mathrm{act}/T)$ \cite{50yKramers,Langer}. Further the average mean velocity of every unit of the chain $v_i=\,2\pi/T_\mathrm{esc}$ is determined by $T_\mathrm{esc}$.

\subsection{First escape scenario} \label{subsec:1stesc}
\begin{figure}
\subfigure[Sketch of probable escape processes \label{fig:escsk}]
{\includegraphics*[bb= 0 -10 200 190,clip,width=0.5\textwidth]{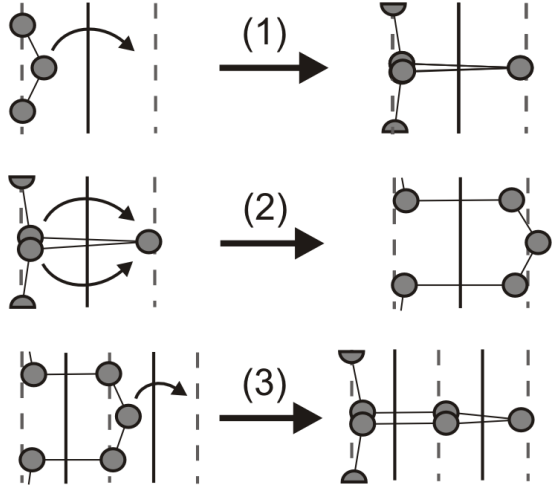}}
\subfigure[Activation energy versus bond length \label{fig:eact}]{\includegraphics[width=0.5\textwidth]{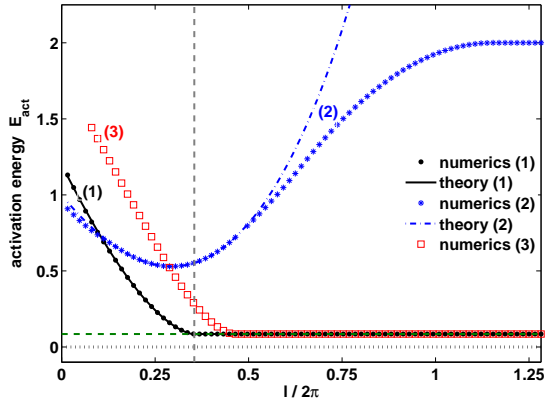}}
\caption{(Color online) Activation energy $E_\mathrm{act}$ as a function of the
bond length $l$. The vertical dashed line represents $x_U^1$
and the horizontal dashed line $E_\mathrm{act}^\mathrm{sat}$ given in
\eq{eq:esat}. The labels relate to the graphs of the activation energy to the corresponding first, second and third escape scenario sketched in (a). Note that the saturation value $E_\mathrm{act}^\mathrm{sat}$ attained by the activation energies $E_\mathrm{act}^{(1)}$ and $E_\mathrm{act}^{(3)}$ in the limit of large bond lengths lies above zero represented by the dotted line. The remaining parameter values are $N=99,n_0=50,D=10,F/F_\mathrm{cr}=0.8$, and $\alpha=0.1$.}
\end{figure}

\noindent We start with the \textit{first escape scenario}. In the latter only the pulled particle escapes from its initial minima to the next well of the on-site potential in positive x-direction while all other units remain close to their starting position. In order to determine the scaling behavior of the activation
energy for the first escape scenario in the limit of weak coupling,
$K \ll 1$, we consider a truncated one-dimensional trimer
model. According to the TSC presented in \figref{fig:profa}, the escape process
involves only three units for $l<x_U^1$. All
other units rest at the minimum of the on-site potential $x_n=0$
under maintenance of the bond length $l$ and hence do not
contribute energy to the activation energy of the escape process.
The condition of stationarity in y-direction $\{\dot{y}_n\}=0$ is fulfilled by setting $y_{n_0}=y_{n_0\pm1}$. Since we observed hairpin-like crossing configurations in the simulations for non-vanishing but sufficiently point force magnitude $F$, we consider only reflection symmetry TSC, i.e. $x_{n_0+1}=x_{n_0-1}$ and $y_{n_0+1}=y_{n_0-1}$. For $l< x_U^1$, we can assume that the stationary TSC is adopted when the pulled unit of the chain is situated close to $\pi$ and the value of the coordinate $x_{n_0\pm1}$ is almost zero. Hence we put
\begin{subequations}
\begin{align}
 x_{n_0}=&\,\pi+\delta x_{n_0}\,, \\
 x_{n_0\pm1}=&\,0+\delta x_{n_0\pm1}\,,
\end{align}
\end{subequations}
into the deterministic part of \eq{eq:dglx} and solve the linearized system of equation $\{\dot{x}_n\}=0$.
Despite that we consider the weak coupling limit, $K\ll 1$, further simplifications of the system of equations by expanding the interaction potential $W(r_{n,n-1})$ up to the $2$nd order is not possible since the bonds are stretched ,$r_{n_0,n_0\pm 1}\gg l$, at the TSC \figref{fig:prof}. Substituting
\begin{align}
A(x)=&\,2D\alpha\left(e^{-\alpha(x-l)}-e^{-2\alpha(x-l)}\right)\,, \label{eq:defA} \\
B(x)=&\,2D\alpha^2\left(e^{-\alpha(x-l)}-2e^{-2\alpha(x-l)}\right)\,, \label{eq:defB}
\end{align}
we get
\begin{subequations}
\begin{align}
 x_{n_0}^{\dagger (1)}=&\,\pi+\frac{4\,B(\pi)F+4\,A(\pi)-2\,F}{1+2\,B(\pi)}\,,\\
 x_{n_0\pm1}^{\dagger (1)}=&\,\frac{4\,B(\pi)F+2\,A(\pi)}{1+2\,B(\pi)}\,.
\end{align} \label{eq:tsc1}
\end{subequations}
For the first escape scenario, the energy of the chain at the
initial metastable state is almost close to $V^{(1)}\left(\{\V{q}^\mathrm{min}\}\right)=0.5\left(1-\sqrt{1-4F^2}-2F\arcsin(2F)\right)$ regardless of the value of the bond length. Using the expressions given in \eq{eq:tsc1}, we obtain the following scaling behavior of the activation energy for the first escape scenario \eq{eq:eact}
\begin{align}
    E_{\mathrm{act}}^{(1)} \propto&\,K\,\left(x_U^1-l-\sqrt{x_U^1-l}\right)\,\left(\mathrm{const}-F\right)\,,\label{eq:eactscale}
\end{align}
for $K \ll 1, l<x_U^1,$ and $F/F_\mathrm{cr}<1$. The numerical results for the activation energy for the first escape scenario and the analytic one are depicted in \figref{fig:eact} and labeled by $(1)$. Both results match very well.

In the limit $l\geq x_U^1$, the transition state
configuration is always given by $x_{n_0}^{\dagger (1)}=x_U^1$ and $x_n^{\dagger (1)}=0\,,\forall
n\neq n_0$, with $r_{n,n+1}^{\dagger (1)}=l$ regardless of the coupling strength. Then the corresponding activation energy attains the saturation value
\begin{align}
 E_{\mathrm{act}}^\mathrm{sat}=\sqrt{1-4F^2}+2F\arcsin(2F)-F\pi\,. \label{eq:esat}
\end{align}
One recognizes that the curve, label $(1)$, of the activation energy in \figref{fig:eact} finally converges to the limit value $E_{\mathrm{act}}^\mathrm{sat}$ represented by the horizontal dashed line.

\subsection{Second escape scenario} \label{subsec:2ndesc}
\noindent In the second escape scenario the $n_{0\pm1}-$th particles escape
from the first well to the next one while all other units remain
close to their starting position. This case is sketched in
\figref{fig:escsk} and labeled by $(2)$. The configuration of the
chain at the initial metastable state can be described by an
effective two particle problem. By using the following ansatz for
the position of the particles
\begin{subequations}
\begin{align}
 x_{n_0}=&\,2\pi+\delta x_{n_0}\,, \\
 x_{n_0\pm1}=&\,0+\delta x_{n_0\pm1}\,.
\end{align}
\end{subequations}
and solving again the linearized stationary system of equation, we get
\begin{subequations}
\begin{align}
 x_{n_0}^{\mathrm{min}(2)}=&\,2\pi+\frac{4\,B(2\pi)F+4\,A(2\pi)-2\,F}{6\,B(2\pi)-1}\,,\\
 x_{n_0\pm1}^{\mathrm{min}(2)}=&\,\frac{4\,B(2\pi)F-2\,A(2\pi)}{6\,B(2\pi)-1}\,.
\end{align} \label{eq:min2}
\end{subequations}
The transition state configuration in the $2$nd escape scenario is determined by an effective three particle
problem. For $l<x_U^2-\pi$, we assume that the stationary
configuration is adopted when the pulled unit is located close to
$2\pi$, its neighbors are located at top of the energy barrier,
situated at $\pi$, and the $n_{0\pm2}-$th particles rest at the
minimum of the starting well. Hence we put
\begin{subequations}
\begin{align}
 x_{n_0}=&\,2\pi+\delta x_{n_0}\,, \\
 x_{n_0\pm1}=&\,\pi+\delta x_{n_0\pm1}\,, \\
 x_{n_0\pm2}=&\,0+\delta x_{n_0\pm2}\,,
\end{align}
\end{subequations}
into the deterministic part of \eq{eq:dgl} and solve again the linearized system of equation under the condition $\{\dot{x}_n\}=\{\dot{y}_n\}=0$. Finally we obtain for the correction terms $\delta x_i$ the solutions
\begin{subequations}
\begin{align}
\delta x_{n_0}^{\dagger (2)}=&\,\frac{(4\,A(\pi)-2F)(1+2\,B(\pi))+8\,B(\pi)^2\,F}{4\,B(\pi)^2+2\,B(\pi)-1}\,,\\
\delta x_{n_0\pm1}^{\dagger (2)}=&\,\frac{2\,B(\pi)\,F+B(\pi)\,\delta x_{n_0}^{\dagger (2)} }{1+2\,B(\pi)}\,, \\
\delta x_{n_0\pm2}^{\dagger (2)}=&\,F-\frac{1}{2}\delta x_{n_0}^{\dagger (2)}+\delta x_{n_0\pm1}^{\dagger (2)}\,.
\end{align} \label{eq:tsc2}
\end{subequations}
In \figref{fig:eact}, label $(2)$, we depict the graphs showing the dependence of the activation
energy on the bond length $l$ for the $2$nd escape process. The theoretical result matches very well with the numerics for $l\lesssim 2\pi$. In particular it has to be emphasized that the position of the minimum of the activation energy as a function of the bond length is very well reproduced by the analytic expression given in \eq{eq:min2} and \eqref{eq:tsc2}. The analytical result deviates from the numerics with further increasing value of the bond length since the theory does not reproduce that the TSC is always given by $x_{n_0}^{\dagger (2)}=x_U^2$, $x_{n_0\pm1}^{\dagger(2)}=\pi$ and $x_{n}^{\dagger (2)}=0$ with $r_{n,n+1}=l$ for $l\geq x_U^2-\pi$. In addition the activation energy eventually attains the saturation value $E_{\mathrm{act}}^{(2)}=2$ in limit of large bond lengths because then the TSC is determined by the given above fixed configuration. 

\subsection{Third escape scenario} \label{subsec:3rdesc}
\noindent At the end of the $2$nd escape process the pulled particle and its
two neighbors are located in the same well of the on-site potential
while all other constituents of the chain remain in the well behind \figref{fig:escsk}.
Therefore the situation for the pulled unit is
similar to the initial condition in the first escape process and the
circumstances for its two neighbors resemble the one at the
beginning of the second escape scenario. Thereupon, either the
$n_{0\pm2}-$th units escape like in the previously mentioned $2$nd
case or the pulled particle escapes further forward. The last case
is sketched in \figref{fig:escsk} and is labeled by $(3)$. The
dependence of activation energy on the bond length is depicted in \figref{fig:eact}. The latter is similar to the results for the $1$st scenario but is evidently shifted to higher energies due to the fact that the  bonds $r_{{n_0\pm1},{n_0\pm2}}$ are stretched during the escape process which is in contrast to the $1$st case. Furthermore it is shown that the activation energy reaches the limit value
$E_{\mathrm{act}}^\mathrm{sat}$ not until $l\gg x_U^1$.

To sum up, we have shown that in the limit of weak coupling, $K \ll
1$, the escape behavior of the chain is governed by consecutive
individual escapes of single oscillators. In the limit of small bond
length $l\ll x_U^1$, the units of the chain escape forward in
$x$-direction by using the escape scenario $(1)$ and $(2)$. For
larger values of the bond length, the $2$nd and $3$rd escape
scenario are utilized by the units in an alternating manner.
Notably the vital second escape scenario, \secref{subsec:2ndesc}, requests a higher amount of
activation energy compared to the other two scenarios and hence governs the time scale of chain transport which is reflected in the mobility.

\section{Mobility} \label{sec:mob}
\noindent Having studied the escape dynamics of the system which is the precondition for transport, in the following we focus our interest on the transport properties of a chain of interacting nonlinear overdamped Brownian particles confined onto a periodic substrate. In the case of a chain driven by a dc point force oriented along a symmetry axis of a $2$D substrate, the stationary transport proceeds in force direction, whereas transverse diffusion is not affected by the bias. When considering interacting Brownian particles, it is appropriate to study the motion of their center of mass (c.o.m.). The position of the c.o.m. in force direction at time $t$ is denoted by $X(t)$. Referring to \eq{eq:dglx}, the LE reads as
\begin{align}
 \dot{X}(t)=\,-\frac{1}{2N}\sum_{i=1}^{N}\sin\left(x_i\right)+\frac{F}{N}+Q(t)\,. \label{eq:dglX}
\end{align}
With the $\delta$-correlated Gaussian white noise $Q(t)=\sum_i^N\xi_i^x(t)/N$. The nonlinear behavior of $X(t)$ results from the term $F_S=\sum_{i=1}^{N}\sin\left(x_i\right)/(2N)$ in \eq{eq:dglX}. The latter can be interpreted as an effective sliding friction force $F_S$ which results from the interaction of the single units with the substrate $U(x_n)$. The motion of the chain is quantitatively assessed by the net velocity of the c.o.m. $X(t)$
\begin{align}
  v_x\equiv&\,\lim_{t\to\infty} \frac{\av{X(t)}}{t}  \label{eq:vx}\,,
\intertext{or, equivalently, by the related mobility}
  \mu_x\equiv&\,\frac{v_x}{F}\,. \label{eq:mob}
\end{align}
In the absence of the on-site potential, the characteristic free net
velocity is $v_x^{\mathrm{free}}=F/N$ which is associated with the mobility of
the chain is $\mu_0=1/N$. Below, the numerically calculated results
for the mobility $\mu_x$ are presented in units of the characteristic
free mobility $\mu_0$ and this is equivalent to express the net
velocity $v_x$ in units of the characteristic free net velocity $v_x^{\mathrm{free}}$, i.e.  $\mu_x/\mu_0=v_x/v_x^{\mathrm{free}}$.

The system of coupled LE  \eq{eq:dgl} has been integrated numerically through a second-order Heun stochastic solver scheme. Starting from a thermal equilibrated configuration in which all constituents of the chain are located near the bottom of one well of $U(x_n)$, the external point force has been applied at the $n_0$-th unit at $t_0=0$. The stochastic trajectories of the constituents have been integrated numerically from $t_0$ with the time step $\Delta t=10^{-2} \ll t_\mathrm{char}$ up to $t_\mathrm{end}=10^{5}$. The characteristic time $t_\mathrm{char}$ is given by the relaxation time for the overdamped motion of one particle in the biased periodic potential $U(x_n)$ in the case of vanishing coupling $t_\mathrm{char}=2/\sqrt{1-4F^2}$. Further average quantities have been obtained as ensemble averages over $100$ trajectories.

\subsection{Role of the bond length} \label{sec:bondlength}
\noindent First we study the influence of the internal degrees of freedom, viz. the bond length $l$, on the mobility $\mu_x$. In \secref{subsec:2ndesc} we showed that in the limit of weak coupling the escape rate $r_\mathrm{esc}$ and the mean first escape time $T_\mathrm{esc}$  of the dominating second escape process, respectively, exhibits a resonance behavior as a function of the bond length. Due to the fact that escape of the units of the chain is necessary for transport of the chain, we expect to observe a characteristic dependence of the velocity of the system on the bond length.

Several authors \cite{Heinsula,mudimer3} have shown that the transport of an underdamped $1$D dimer system in a periodic potential strongly depends on its bond length $l$. In particular, it was found that the mobility $\mu_x(l)$ is a reflection symmetric function which attains its maximum value at $l\approx (2\,k+1)\pi\,,\,k\in\Z,$ and varies periodically with $\mathrm{mod(2\pi)}$ regardless of the coupling strength and force magnitude $F$ \cite{Heinsula,mudimer3}.

Interestingly, it turns out that the mobility of the $2$D discrete nonlinear coupled oscillator chain shows a resonance behavior as a function of the bond length $l$ with one single maximum within a period of the substrate potential. This means there exist a set of optimal parameter values maximizing the mobility and optimizing the transport properties of the system, respectively. Further the mobility as a function of system parameter exhibits properties which coincide with the results presented in \cite{Heinsula,mudimer3} but also shows several new phenomena.

\begin{figure}
\subfigure[ \label{fig:mobAl}]
{\includegraphics[width=0.5\textwidth]{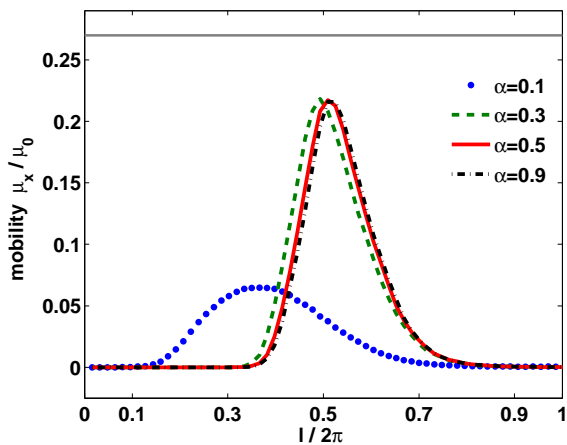}}
\subfigure[ \label{fig:mobAr}]
{\includegraphics[width=0.5\textwidth]{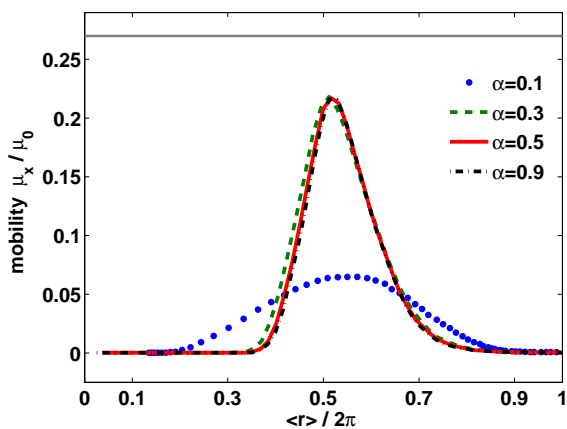}}
\caption{(Color online) C.o.m. mobility versus the bond length $l$ (a) and the
mobility versus the mean distance $\av{r}$ (b) for different values of the inverse interaction range $\alpha$ as indicated in the plot and with the relation in \eq{eq:Kdef} the values of the assigned coupling strength $K$ are $0.4,3.6,10,$ and $32.4$. The horizontal solid line represents the value of  $\mu_x$ of one single overdamped Brownian particle. The remaining parameter values are $N=9,n_0=5,D=10,F/F_{\mathrm{cr}}=0.8$, and $T=0.1$. \label{fig:mobAlsub}}
\end{figure}

Let us first have a look on the influence of the coupling strength $K$ on the mobility as a function of the bond length. In \figref{fig:mobAl} the numerical results for the mobility versus the bond length $l$ for various values of $\alpha$ are depicted. We remind that the inverse interaction range $\alpha$ and the coupling strength $K$ are related according to \eq{eq:Kdef}. It turns out that there exist a finite interval of the bond lengths, $l_c^{\mathrm{low}}<l<l_c^{\mathrm{up}}$, in which translocation occurs. Further it is recognizable that with increasing coupling strength $K$ the lower limit $l_c^{\mathrm{low}}$ approaches from below the position of the unstable state of the on-site potential $x_U^1$, see \eq{eq:stateU}. This is in compliance with the results presented in \secref{subsec:1stesc} since for $l\geq x_U^1$ the activation energy of the first escape scenario is determined by the saturation value $E_\mathrm{act}^\mathrm{sat}$ \eq{eq:esat} which is independent of the coupling strength $K$. In contrast in the range $l<x_U^1$, the number of units participating in the first escape event, and therewith connected the activation energy, grows with increasing coupling strength. Hence the mean first escape time $T_\mathrm{esc} \propto \exp(E_\mathrm{act}^{(1)}/T)$ increases with the value of $K$ and thus the mobility goes to zero in the limit $K\to \infty$. The upper limit $l_c^{\mathrm{up}}$ turns out to be independent of the coupling strength. Hence the larger the value of $K$ the narrower is the interval of the bond lengths where the chain is mobile for a given $F$ and $T$.

In addition, it is shown that the mobility grows monotonically with increasing value of $l$ from $l_c^{\mathrm{low}}$ till the bond length is equal to $l_{\mathrm{peak}}$, where the mobility possesses its maximal value $\mu_x^\mathrm{max}$. With upon growing bond length, $l>l_{\mathrm{peak}}$, the mobility decays till $l\leq l_c^{\mathrm{up}}$.  It turns out that the value of $l_{\mathrm{peak}}$ depends on the coupling strength $K$. In the case of weak coupling, $K<1$, the position $l_{\mathrm{peak}}$ is left from $\pi$ which is in contrast to the results obtained for the 1D underdamped dimer \cite{Heinsula,mudimer3}. With increasing coupling strength $K$ the position $l_{\mathrm{peak}}$ approaches from below $\pi$. 

Further it is demonstrated that the peak height of the mobility $\mu_x^\mathrm{max}$ increases with the coupling strength till $K$ reaches the value $K_{\mathrm{cr}}$. For $K>K_{\mathrm{cr}}$, the maximum value of the mobility remains almost constant upon changing $K$ for a given value of $F$ and $T$. Note that the shape of the mobility graph is asymmetric with respect to $l_{\mathrm{peak}}$ which is in contrast to the results obtained for the 1D underdamped dimer \cite{Heinsula,mudimer3}.

Furthermore, the mobility of one single overdamped Brownian particle can be calculated by means of the Stratonovich formula \cite{Stratonovich,Difftilted}. Comparing the mobility of a monomer, represented by the horizontal solid line in \figref{fig:mobAl}, with the presented maximum values of $\mu_x(l)$, one concludes that the response of the extended $2$D chain to the external driving is less than the one of the monomer. But in the limit of strong coupling both responses nearly coincide. This indicates that the chain consisting of nonlinear coupled oscillator synchronizes and thus behaves like one heavy Brownian particle with mass $N$.

Due to the mutual impact of stochastic forces, the point
force and the interaction with the substrate  the bonds of the chain experience  dynamical alterations. Therefore the adopted distance between two adjacent units mostly differs from the bond length $l$ during the motion of the chain. To gain more insight, we discuss the mobility versus the
mean distance $\av{r}$ between two neighboring units in the long time limit averaged over all
sites $\av{r}=\,\lim_{t \to \infty} \sum_{i=1}^{N-1}\av{r_{i,i+1}(t)}/(N-1)$. In \figref{fig:mobAr} the numerical results for the c.o.m. mobility $\mu_x$ versus $\av{r}$ for various coupling strengths are presented. In contrast to the results shown in \figref{fig:mobAl}, the mobility always reaches the maximal value at $\av{r}_{\mathrm{peak}}=\pi$ regardless of the value of $K$. Assuming that the position of a particle $x_{i+1}=x_{i}+\av{r}$ can be described by means of the averaged distance between two units, one finds that for $\av{r}=\pi$ the effective sliding friction force $F_S$, see \eq{eq:dglX}, possesses its minimum value and thus the mobility reaches its maximum value.

\begin{figure}[t]
\includegraphics[width=0.5\textwidth]{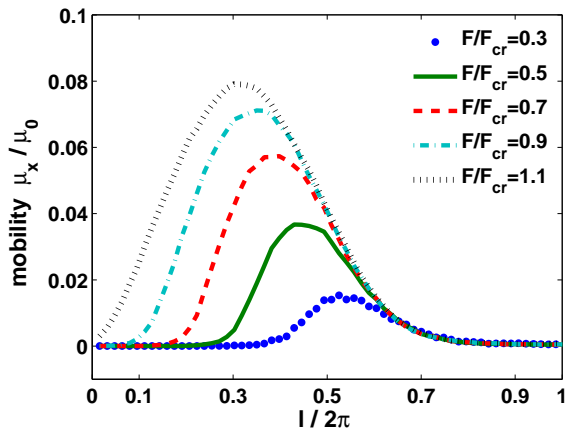}
\caption{(Color online) C.o.m. mobility versus the bond length for different values
of the point force magnitude $F$. Parameters values are $N=9,n_0=5,D=10,\alpha=0.1$,\- and
$T=0.1$. \label{fig:mobFsub}}
\end{figure}

Subsequently we study the influence of the point force magnitude $F$ on the
mobility $\mu_x$ for given coupling strength and temperature. The results are presented
in \figref{fig:mobFsub}. It is shown that the value of the lower limit value $l_c^{\mathrm{low}}$ is strongly influenced by the point force magnitude $F$, more precisely the larger the value of $F$ the less is the value of $l_c^{\mathrm{low}}$. Referring to the scaling behavior of $E_\mathrm{act}^{(1)}$ given in \eq{eq:eactscale}, the activation energy of the first escape process decreases linearly with the value of $F$ for a given bond length $l$. Hence, the lower limit value $l_c^{\mathrm{low}}$ goes to zero for $F/F_\mathrm{cr} \to 1$. In contrast, it turns out that the upper limit value $l_c^{\mathrm{up}}$ is not influenced by the point force magnitude $F$. Thus the region where the mobility differs significantly from zero strongly depends on $F$. Similar to the previously discussed results, we find that the adopted mean distance at the maximum value of the mobility $\av{r}_{\mathrm{peak}}$ is equal to $\pi$. (The corresponding panel is not presented.) Since the deviation of the bond length $l$ from $\av{r}$ grows with increasing value of $F$, one notices that the value of $l_{\mathrm{peak}}$ gets smaller when enlarging the point force magnitudes $F$. In addition the corresponding peak height increases monotonically with the point force magnitude $F$.

\begin{figure}[t]
 \includegraphics[width=0.5\textwidth]{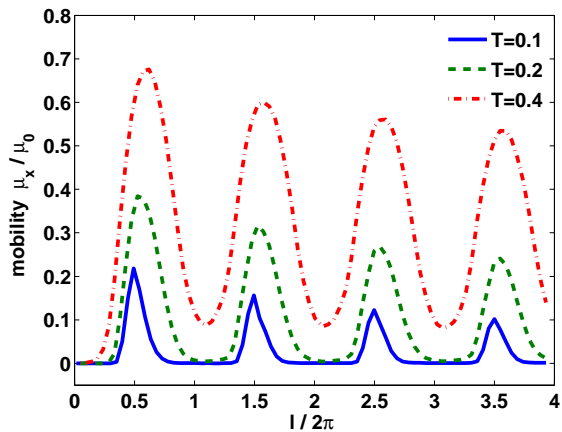}
\caption{(Color online) C.o.m mobility versus the bond length $l$ for different values of the
temperature $T$. The remaining parameter values are $N=9,n_0=5,D=10,\alpha=0.5$, and $F/F_\mathrm{cr}=0.8$. \label{fig:mobperi}}
\end{figure}

Finally, we present the results for the impact of the temperature $T$ on the mobility in \figref{fig:mobperi}. For low temperature, ratios of $l/2\pi$ exist where escape and therefore translocation is impossible. Upon increasing $T$, the region around $l\approx
\pi$, where the mobility of the chain is significantly different
from zero, becomes broader. Eventually for high enough $T$, the chain is considerably
mobile for all values of $l$. In addition, it is illustrated that the peak
height at $l_{\mathrm{peak}}$ increases monotonically with $T$, remaining always smaller than $\mu_0$.

In \figref{fig:mobperi} the influence of the bond length on the mobility for a broader interval of values of $l$ is depicted. It is shown that $\mu_x$ varies periodically with $\mathrm{mod}(2\pi)$. As a novel feature it is found that the peak height at $l_\mathrm{peak}$ decreases upon increasing bond length $l$. This effect is independent of the value of the remaining system parameters. As pointed out in \secref{sec:eact}, the sequence of occurring escape processes depends on the value of $l$. For $l+n2\pi$, $l\in [0,2\pi]$, the escape rate $r_\mathrm{esc}$ of the first $n$ escape processes is determined by $r_\mathrm{esc} \propto \exp(-E_\mathrm{act}^\mathrm{sat}/T)$ where $E_\mathrm{act}^\mathrm{sat}$ is given in \eq{eq:esat}. After the time $t_1=n\,T \approx n\,\exp(E_\mathrm{act}^\mathrm{sat}/T)$ the subsequent escape processes are similar to the first and second escape scenario respectively for $l \in [0,2\pi]$. Then the system needs the time $t_2 \approx n\,(\exp(E_\mathrm{act}^{(1)}(l)/T)+\exp(E_\mathrm{act}^{(2)}(l)/T))$ till the next escape process occurs similar to the one that has taken place before for $l \in [0,2\pi]$. Since the mean time till the oscillator escapes, utilizing the $2$nd escape scenario, increases with the bond length $l$ the peak height at $l_\mathrm{peak}$ decreases with the latter. 

\subsection{Locked-running transition} \label{sec:locked}
\noindent Now we study the impact of the point force magnitude $F$ on the
mobility for a fixed value of the bond length. In the case of one
single overdamped Brownian particle, $\mu_x$ and $F$ are connected
by the Stratonovich formula \cite{Stratonovich,Difftilted}. A
locked-running transition of the system occurs if $F>F_d$ with the
depinning force $F_d$. The latter is defined in such a way that a
small change of the external driving strength $F=F_d+\varepsilon$,
$\varepsilon \ll 1$, results in a significant enhancement of the
mobility.

\begin{figure}
\includegraphics[width=0.5\textwidth]{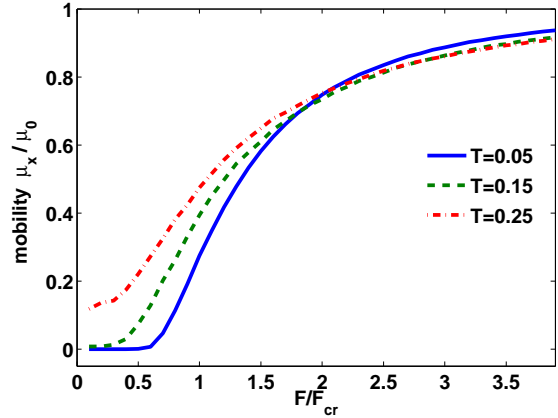}
\caption{(Color online) C.o.m. mobility versus point force magnitude $F$ for different temperatures $T$. The remaining parameters values are $N=9,n_0=5,D=10,\alpha=0.5$,\- and
$l=3.1$. \label{fig:mobFT}}
\end{figure}

In \figref{fig:mobFT} the dependence of $\mu_x$ on $F$ is presented
for different temperatures $T$. The value of the bond length is
fixed at $l=3.1$ a value for which the chain is considerably mobile
regardless of the coupling strength $K$. In general, since one cannot find a
preferential direction of the random Brownian motion in a spatially periodic potential at
thermal equilibrium \cite{Reimann02} the velocity $v_x$ attains the smallest value for $F=0$
independent of the value of $T$. The occurrence of a directed motion $v_x \neq 0$ would be in contradiction to the second law of thermodynamics. Note that for $F=0$ the mobility is not defined, see \eq{eq:mob}. With further increasing value of $F$ the mobility grows monotonically and finally goes to $\mu_0$ in the
limit $F\to \infty$. 

From \figref{fig:mobFT} we deduce that the
value of the depinning force $F_d$ goes to zero with increasing
temperature $T$. In particular for $T=0.25$, the mobility attains a
non-zero value regardless of $F$. For a sufficiently weak point force magnitude, $F/F_\mathrm{cr}<1$, the mobility increases for stronger thermal fluctuations, that is higher
temperatures. In this parameter region the motion of the chain is
mainly instigated by the thermal fluctuations. Additionally, it turns out that in the limit of overcritical external driving, $F/F_\mathrm{cr} \gg 1$, the slope of the mobility as a function of the point force decreases for higher temperatures, i.e. we observe a
\textit{noise-induced suppression of the mobility}.

Since the escape rate and therewith connected the mobility increases with growing
value of the temperature according to the Arrhenius-Van't Hoff law, the observed result seems to be counter-intuitive. By applying the point force at the $n_0$-th unit, its effective energy landscape
changes. The motion of the particles, which are close to the
pulled one, proceeds preferably in point force direction due to the lower energy barrier. The
situation for units located nearby the
endings of the chain is different. Due to the symmetry of the on-site potential the
probabilities, respectively, the rates to escape forward or backward are
equal. Hence it is possible that single units or a segment of the
chain escape in direction opposite to the one determined by the
point force. Due to the fact that the mean time which the
center of mass needs to move forward a certain distance grows with
increasing value of $T$, the effective c.o.m. velocity in force direction becomes lower.

\begin{figure}
\includegraphics[width=0.5\textwidth]{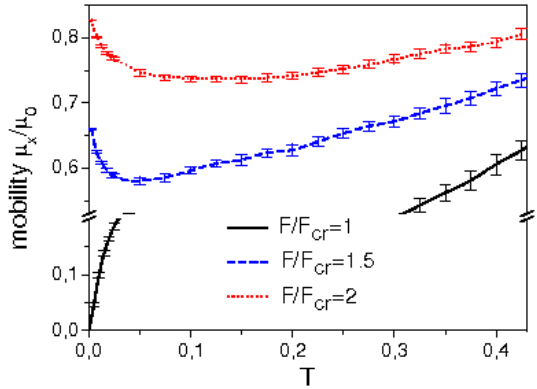}
\caption{(Color online) C.o.m. mobility versus temperatures $T$ for various point force magnitudes $F$. For the sake of clarity there is a break in the y-axes. Note that the curve for $F/F_\mathrm{cr}=1$, represented by the solid line, is a monotonically increasing function of the temperature $T$. The remaining parameter values are $N=9,n_0=5,D=10,\alpha=0.5$,\- and $l=3.1$. \label{fig:mobneg}}
\end{figure}

Finally we study the dependence of
the mobility on temperature for certain fixed point force magnitudes
$F$. The results are presented in \figref{fig:mobneg}. In the limit
of small values of $F$, $F/F_\mathrm{cr}\leq 1$, the motion of the chain is instigated by thermal fluctuations and thus the c.o.m. mobility $\mu_x$ strongly depends on the temperature $T$. For $T=0$ the mobility is equal to zero. With further increasing value of the temperature $T$ the mobility $\mu_x$ grows slightly linear. In contrast in the limit of a sufficiently strong
external driving, $F/F_\mathrm{cr}\gg 1$, the motion of the chain is purely
induced by the external point force $F$ and thus $\mu_x \neq 0$ for
$T=0$. Upon increasing the temperature the mobility decreases and
finally reaches a minimum at a finite temperature. By further enhancing the strength of the thermal fluctuation $T$, $\mu_x$ grows slightly linear regardless of the point
force magnitude. This phenomenon is called \textit{negative resistance} \cite{Bao,Borromeo,Cecchi}.
Comparing the curves for $F/F_\mathrm{cr}=1.5$ and $F/F_\mathrm{cr}=2$, it turns out that the value of the critical temperature depends on the point force magnitude, more precisely, the latter is shifted to
higher values for stronger point force magnitudes $F$. 


\section{Summary}
\noindent In summary, we have considered the thermal activated motion of an extended
two-dimensional discrete oscillator chain absorbed on a periodically structured
substrate under the influence of a localized point force. Attention
has been paid to the escape dynamics of the chain from the
metastable states of the substrate potential which is accomplished by the adaption of kink-like excitations -- also referred to as the transition state. The shape of the latter and the
corresponding activation energies have been calculated. Due
to the fact that the bending rigidity is equal to zero, the obtained
transition structures resemble the shape of very thin needles.
As a novel feature we have found that the transport of the chain in point force direction is determined by stepwise escapes of a single unit or segments of the chain due to the existence of multiple locally stable attractors. In the limit of strong coupling, $K \gg 1$, the units exhibit an organized
collective behavior and the chain escapes like a rigid unit from one domain of attraction to the subsequent one. In contrast for weak coupling, $K \ll 1$, it has been found that the escape behavior of the chain is governed by consecutive individual escape steps of single oscillators. Therefore, we have identified the first three possible escape scenarios with the lowest amount of activation energy.
Interestingly, it has been shown that the vital second escape scenario, which requests a higher amount of activation energy compared to the other two scenarios and thus
governs the time scale it takes for the chain to be transported by
one period of the substrate potential, possesses a minimum at a certain
value of the bond length $l$. From this follows that for unfavorable
ratios of the bond length the escape will be highly improbable while
other ratios will bear good conditions for the chain to escape. The
complicated dependence of the activation energy, respectively, of
the escape rate on the bond length is reflected in a non-monotonous
behavior of the center of mass mobility. The latter exhibits
features which previously have been found in models considering one harmonically coupled one-dimensional dimer \cite{Heinsula,mudimer3}, viz. the mobility reflects the
periodicity of the substrate potential and the latter reaches its
maximum value if and only if the mean distance between the two coupled
units equals an odd integer number times the value of the half
periodicity of the periodic potential. In comparison we have
presented several new transport phenomena. For sufficiently weak
external driving, $F/F_\mathrm{cr}<1$, the mobility shows a resonance behavior as a function of the bond length $l$ with one single maximum within each period of the substrate potential whose position
$l_\mathrm{peak}$ depends on all system parameters. In general it turned out that the mobility attains its maximum value if and only if the mean distance $\av{r}$ between the coupled units equals $\av{r}_\mathrm{peak}=(2k+1)\pi, k \in \Z$. The connection between $l_\mathrm{peak}$ and $\av{r}_\mathrm{peak}$ is determined by the interplay of different acting forces. Further we have shown that the maximal value of the mobility $\mu_x^\mathrm{max}$ at $l_\mathrm{peak}$ decays
upon increasing bond length. Lastly, the role of the thermal
fluctuations and the external driving played for the activated motion of the
chain has been considered. While it has been found that for
sufficiently weak point force magnitude the mobility grows
monotonically with increasing temperature, for a given overcritical
external driving force the phenomenon of negative resistance has been found, i.e. the mobility possesses a minimum at a finite value of the strength of the thermal fluctuations.

\vspace*{0.25cm}

\centerline{\large{\bf Acknowledgments}}

\noindent This research was supported by SFB 555 (L.S.-G., S.F.) and, as well, by the VW
Foundation Project I/80425 (L.S.-G., S.M.).


\end{document}